\documentclass[graybox]{svmult}

\usepackage{type1cm} 
\usepackage{makeidx} 
\usepackage{graphicx} 
\usepackage{multicol} 
\usepackage[bottom]{footmisc}

\usepackage{amsmath}
\usepackage{amsfonts}
\usepackage{amssymb}
\usepackage{braket}
\usepackage{placeins}
\usepackage{xcolor}
\usepackage{soul}
\usepackage{aastex_macro}
\usepackage{multirow}
\usepackage{newtxtext} %
\usepackage{newtxmath} 
\usepackage[square,numbers]{natbib}

\makeindex 

\begin{document}

\title*{Rejection criteria based on outliers in the KiDS photometric redshifts and PDF distributions derived by machine learning }
\titlerunning{Rejection criteria for zphot PDFs}  

\author{Valeria Amaro, Stefano Cavuoti, Massimo Brescia, Giuseppe Riccio, Crescenzo Tortora, Maurizio D'Addona, Michele Delli Veneri, Nicola R. Napolitano, Mario Radovich, Giuseppe Longo} 

\authorrunning{Amaro et al. 2020}
\institute{Valeria Amaro and Nicola R. Napolitano \at School of Physics and Astronomy, Sun Yat-sen University, Zhuhai Campus, Guangzhou 519082, PR China \email{valeriaa@mail.sysu.edu.cn}
\and Massimo Brescia, Stefano Cavuoti, Giuseppe Riccio and Crescenzo Tortora \at INAF - Astronomical Observatory of Capodimonte, Salita Moiariello 16, I-80131 Napoli, Italy. \email{stefano.cavuoti@gmail.com}
\and Maurizio D'Addona and Giuseppe Longo \at Department of Physics, University of Naples Federico II, Strada Vicinale Cupa Cintia, 21, I-80126 Napoli, Italy. \email{longo@na.infn.it} 
\and Michele Delli Veneri \at DIETI, University of Naples Federico II, via Claudio 21, I-80125 Napoli, Italy \email{micheledelliveneri@gmail.com}
\and Mario Radovich \at INAF – Osservatorio Astronomico di Padova, Vicolo Osservatorio 5, 35122 - Padova, Italy 
\email{mario.radovich@inaf.it}}

%
\maketitle
\textit{Preprint version of the manuscript to appear in the Volume "Intelligent Astrophysics" of the series "Emergence, Complexity and Computation", Book eds. I. Zelinka, D. Baron, M. Brescia, Springer Nature Switzerland, ISSN: 2194-7287} \\

\abstract{
The Probability Density Function (PDF) provides an estimate of the photometric redshift (zphot) prediction error. It is crucial for current and future sky surveys, characterized by strict requirements on the zphot precision, reliability and completeness. 
The present work stands on the assumption that properly defined rejection criteria, capable of identifying and rejecting potential outliers, can increase the precision of zphot estimates and of their cumulative PDF, without sacrificing much in terms of completeness of the sample.  We provide a way to assess rejection through proper cuts on the shape descriptors of a PDF, such as the width and the height of the maximum PDF's peak. In this work we tested these rejection criteria to galaxies with photometry extracted from the Kilo Degree Survey (KiDS) ESO Data Release 4, proving that such approach could lead to significant improvements to the zphot quality: e.g., for the clipped sample showing the best trade-off between precision and completeness, we achieve a reduction in outliers fraction of $\simeq 75\%$ and an improvement of $\simeq 6\%$ for NMAD, with respect to the original data set, preserving the $\simeq 93\%$ of its content.}

\abstract*{
The Probability Density Function (PDF) provides an estimate of the photometric redshift (zphot) prediction error. It is crucial for current and future sky surveys, characterized by strict requirements on the zphot precision, reliability and completeness. 
The present work stands on the assumption that properly defined rejection criteria, capable of identifying and rejecting potential outliers, can increase the precision of zphot estimates and of their cumulative PDF, without sacrificing much in terms of completeness of the sample.  We provide a way to assess rejection through proper cuts on the shape descriptors of a PDF, such as the width and the height of the maximum PDF's peak. In this work we tested these rejection criteria to galaxies with photometry extracted from the Kilo Degree Survey (KiDS) ESO Data Release 4, proving that such approach could lead to significant improvements to the zphot quality: e.g., for the clipped sample showing the best trade-off between precision and completeness, we achieve a reduction in outliers fraction of $\simeq 75\%$ and an improvement of $\simeq 6\%$ for NMAD, with respect to the original data set, preserving the $\simeq 93\%$ of its content.}


\section{Introduction}
\label{sec:intro}
Photometric redshifts (zphot) are crucial for modern cosmology surveys, since they provide the only viable approach to determine the distances of large samples of galaxies. Over the years they have been used to constrain the dark matter and dark energy contents of the Universe through weak gravitational lensing \cite{Serjeant2014,Hildebrandt2017,Fu2018}, to reconstruct the cosmic Large Scale Structure \cite{Aragon2015}, to identify galaxy clusters and groups \cite{Capozzi2009,Annunziatella2016,Rad2017}, to disentangle the nature of astronomical sources \cite{Brescia2012,Tortora2016}; to map the galaxy colour-redshift relationships \cite{Masters2015} and to measure the baryonic acoustic oscillations spectrum \cite{Gorecki2014,Ross2017}.

Baum \cite{Baum1962} first  noticed that the stretching of a galaxy spectrum due to the redshift affects the observed colours and hence, if the correlation between photometry and redshift can be uncovered, multi-band photometry could become a powerful tool to estimate redshifts.\\
However it became immediately apparent that such correlation is highly non-linear and too complex to be derived analytically \cite{Connolly1995} and that the derivation of zphot required alternative, interpolative approaches. \\
\indent Nowadays, it is common praxis to divide these methods into two broad classes: the Spectral Energy Distribution (SED) template fitting methods (e.g., \cite{Bolzonella2000, Arnouts1999, Tanaka2015}) and the empirical (or interpolative) methods (e.g., \cite{Tagliaferri2002,Firth2003,Ball2008,CeB2013,Brescia2014b,Graff2014,Cavuoti2015a,Cavuoti2015,Sadeh2016,Soo2018,DIsanto2018}), both characterized by their pros and cons. \\
\indent SED methods rely on fitting the multi-wavelength photometric observations of the objects to a library of synthetic or observed template SEDs, shifted to create synthetic magnitudes for each galaxy template as a function of the redshift. SED fitting methods, while relying on many assumptions, allow pushing zphot beyond the spectroscopic limit.\\
\indent Empirical methods use instead an a priori spectroscopic knowledge (zspec) for a subsample of objects to infer the complicated relationship existing between the photometric data (i.e. magnitudes and or derived colours, in some cases complemented by morphological information) and the redshift. Among these methods, a critical role is played by machine learning (ML).
Among the many ML models applied to the zphot estimation, we quote just a few: neural networks, boosted decision trees, random forests, self-organized maps, convolutional neural networks (see \cite{Fluke2020} and references therein). 
A primary advantage of ML is the high accuracy of predicted zphot within limits imposed by the spectroscopic knowledge base (KB). On the other hand, ML methods have an inferior capability to extrapolate information outside the regions of the parameter space properly sampled by the training data and, for instance, they cannot be used to estimate the redshift of objects fainter than those present in the spectroscopic sample. \\
\indent Extensive reviews of both approaches can be found in \cite{Hildebrandt2010,Abdalla2011,Sanchez2014}.
An additional difference between the two approaches is that SED fitting methods allow obtaining, at once, the zphot, the spectral type of the objects and the Probability Density Function (hereafter, PDF) of the predicted zphot. In contrast, ML-based methods do not naturally provide a PDF
unless special procedures are implemented.\\
The essential complementarity of the two methodologies was proven to be the most reliable and efficient way to produce a high-quality zphot catalogue \cite{Cavuoti2017b}, particularly suitable for extensive surveys, like Euclid (Desprez et al., in prep.) and VRST \cite{Schmidt2020}. 
In this work, we explore the possibility to improve the quality of zphot predictions by excluding from the data potential outliers without loosing much in completeness.\\
The work is structured as follows: in Sec.~\ref{sec:PDF}, we introduce some aspects of PDF evaluation; in Sec.~\ref{sec:data}, we describe the photometry and spectroscopy used for the analysis herein; in Sec.~\ref{sec:method}, we give a description of the method for the rejection criteria identification as well as of the statistical estimators used to quantify the performance of both zphot and cumulative PDF statistics. In Sec.~\ref{sec:results}, we show all the results, and, finally, in Sec.~\ref{sec:conclusion}, we draw the conclusions.

\section{Probability Density Function}\label{sec:PDF}

In general terms, a PDF is a way to parametrise the uncertainty on the zphot prediction and to provide a robust estimate of the reliability of any individual redshift. From a rigorous statistical point of view, however, a PDF is an intrinsic property of a particular phenomenon, regardless of the measurement methods that allow quantifying the phenomenon itself \cite{Brescia2018}.
 
Unfortunately, in the zphot context, the PDF depends both on the measurement methods (and chosen internal parameters of the methods themselves) as well as on the underlying physical assumptions. In this sense, the definition of a PDF in the context of zphot estimation needs to be taken with some caution \cite{Amaro}. \\
\indent The factors affecting the reliability of zphot PDFs are: photometric errors, intrinsic errors of the methods and statistical biases. The PDF becomes, therefore, just a way to somehow compress the information contained in a single error estimate. In other words, the parametrization of a single error, through a probability, allows to cover an entire redshift range (with the chosen bin accuracy), thus leading to an increase of the information rate in order to match the precision required by a specific scientific goal (cf. for instance, the cases of the determination of cosmological parameters \cite{Mandelbaum2008}, and weak lensing measurements \cite{Viola2015}). \\
Therefore, over the last few years, much attention has been paid to develop methods able to compute a full zphot PDF for both individual sources and entire galaxy samples \cite{brammer2008,Ilbert2006,benitez2000,Bonnett2015,CeB2013a,carrasco2014a,carrasco2014b}.\\
The study of the PDFs and their properties (see Sec.~\ref{sec:method}) represents a useful tool to test zphot reliability. In fact, rejection criteria, aimed at removing unreliable zphot and PDFs estimates, as long as they are reproducible in the photometric space, can improve the precision of the results in terms of both NMAD and fraction of outliers for zphot estimates, as well as  the quality of individual PDFs and their cumulative performances described by the statistical indicators discussed in Sec.~\ref{sec:method}. Of course, this comes at the cost of a decreased completeness.

\section{Data}\label{sec:data}

\begin{figure*}
 \centering
 {\includegraphics[width=0.8 \textwidth]{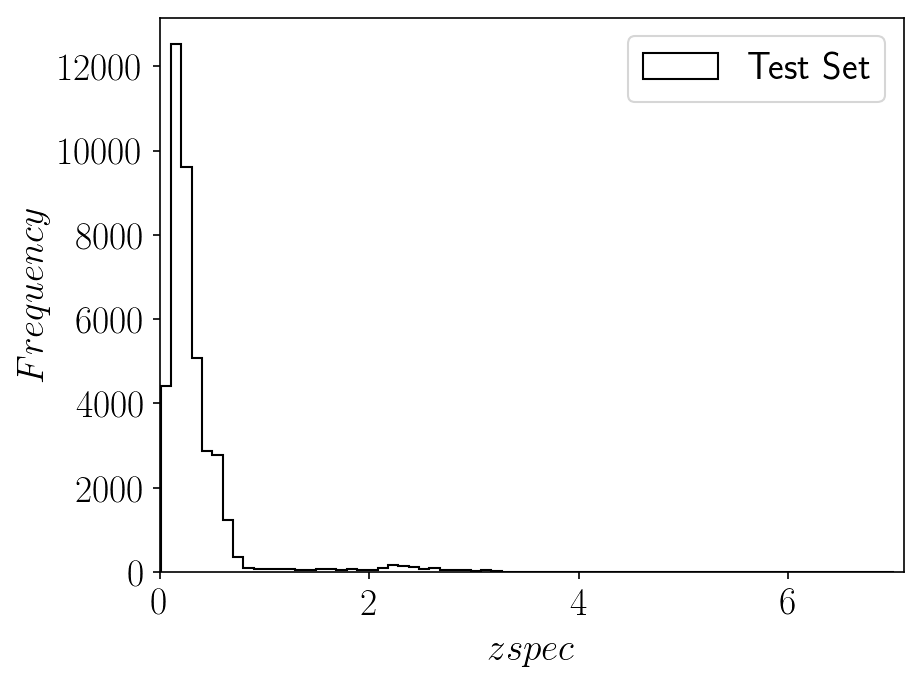}}
 {\includegraphics[width=0.8 \textwidth]{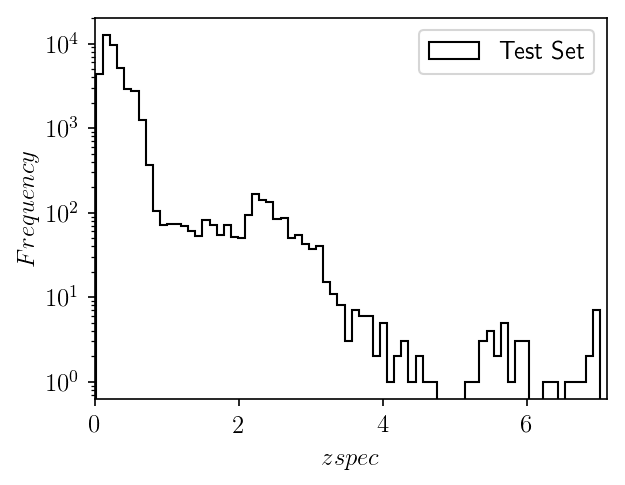}}
 \caption{The zspec distribution of the initial test set ($40,796$ sources). The bottom diagram is the same top plot but in a logarithmic scale.}
\label{fig:zspec_testset_distrib}
\end{figure*}

As spectroscopic knowledge base we used zspec for $136,057$ galaxies extracted from the fourth Data Release (DR) of the ESO Public Kilo-Degree Survey (hereafter, KiDS-ESO-DR4, \cite{Kuijken2019}) which combines data from KiDS and the VISTA Kilo degree INfrared Galaxy survey (VIKING; \cite{Edge2013}). \\
\indent KiDS is an optical survey of about $1350$ $deg^{2}$ carried in 4 bands (ugri) with limiting magnitude r=25.0 AB (5 $\sigma$ in 2''), i.e. 2.5 magnitudes deeper than the Sloan Digital Sky Survey (SDSS), in good seeing conditions ($\sim$0.7'' median full width at half-maximum, FWHM, in the r band). KiDS has been complemented with the NIR photometry in the five bands Z, Y, J, H and Ks from the VIKING survey.\\
The survey is complemented by a set of spectroscopic observations available within the KiDS collaboration as well as from other surveys: COSMOS \cite {Davies2015}, zCOSMOS \cite {Lilly2009}, CDFS \cite {Cooper2012}, DEEP2 \cite{Newman2013} and GAMA DR2 and DR3 \cite {Liske2015,Baldry2018} fields public data.\\
The photometry used in this work consists of the $9$ Gaussian Aperture and PSF (GAaP) magnitudes (u, g, r, i, Z, Y, J, H, Ks), corrected for extinction and zero-point oﬀsets, and 8 derived colours, for a total of 17 photometric parameters for each object. After investigating the photometric parameter distribution for the sample, the data was cleaned by cutting the tails of the magnitude distributions in order to ensure a homogeneous distribution of training points in the parameter space. \\
The spectroscopic data set has been randomly shuffled and split in a $70\%$ training set and a $30\%$ test set ($95,261$ and $40,796$ sources, respectively). We stress that objects in the test set used to evaluate and validate  the trained model are not used during the training phase of the methods (see Sec.~\ref{sec:method}).
In what follows, we used PDFs obtained using METAPHOR (Machine-learning Estimation Tool for Accurate PHOtometric Redshifts \cite{cavuoti2017}) on KiDS-ESO-DR4 data (cf. \cite{Amaro} for details).\\
METAPHOR is a modular workflow designed to produce both zphot and related PDFs. The internal zphot estimation engine is MLPQNA (Multi-Layer Perceptron trained with Quasi-Newton Algorithm; \cite{Brescia2013,Brescia2014a}) while for the zphot we used the \textit{best-estimates} as defined in \cite{Amaro}.\\
We just recall that the selected binning step in zphot used to estimate individual PDFs is $0.02$ and the spectroscopic depth available is equal to zspec=7.01. The zspec distribution for the whole test set is shown in Fig.~\ref{fig:zspec_testset_distrib}.

\section{Methods}\label{sec:method}
\indent The zphot statistics are calculated on the residuals:
\begin{equation}\label{equationDelta}
\Delta z = (z_{spec}-z_{phot})/(1+z_{spec})
\end{equation}
using as zphot the \textit{best-estimate} referenced in Sec.~\ref{sec:data}.\\
We use as accuracy estimators the mean (or bias), the fraction of catastrophic outliers, defined as those objects for which $|\Delta z| > 0.15$, and the normalized median absolute deviation (NMAD), defined as:
\begin{equation}\label{equationNMAD}
NMAD = 1.4826 \times \textit{median} (|\Delta z - \textit{median}(\Delta z)|)
\end{equation}
\indent The shapes of the PDFs, and therefore their intrinsic quality, can be characterized in terms of:
 \begin{itemize}
 \item \textit{pdfWidth}: the width of the PDF in terms of redshift;
 \item \textit{pdfNBins}: the total number of bins of chosen amplitude (which defines the accuracy of the PDF itself), in which the PDF is different from $0$; \item \textit{pdfPeakHeight}: the amplitude of the peak of the PDF, i.e. the value of the maximum probability of the PDF.
 \end{itemize} 
 The cumulative performance of the stacked PDF on the entire sample is instead evaluated by means of the following three estimators:
\begin{itemize}
\item $f_{0.05}$: the percentage of residuals $\Delta$z within $\pm 0.05$;
\item $f_{0.15}$: the percentage of residuals $\Delta$z within $\pm 0.15$;
\item $\Braket{\Delta z}$: the average of all the residuals $\Delta$z of the stacked PDFs.
\end{itemize}
\noindent where by stacked PDFs we mean the individual zphot PDFs transformed into the PDFs of scaled residuals $\Delta z$  and then stacked for the entire sample.\\

\indent Furthermore, the quality of the individual PDFs is evaluated against the single corresponding zspec in the test set, by defining five types of occurrences:
\begin{itemize}
\item \textit{zspecClass} = 0: the zspec is within the \textit{bin} containing the peak of the PDF;
\item \textit{zspecClass} = 1: the zspec falls in one bin from the peak of the PDF;
\item \textit{zspecClass} = 2: the zspec falls into the PDF, e.g. in a bin in which the PDF is different from zero;
\item \textit{zspecClass} = 3: the zspec falls in the first bin outside the limits of the PDF;
\item \textit{zspecClass} = 4: the zspec falls out of the first bin outside the limits of the PDF.
\end{itemize} 

\begin{figure*}
 \centering
 {\includegraphics[width=0.7 \textwidth]{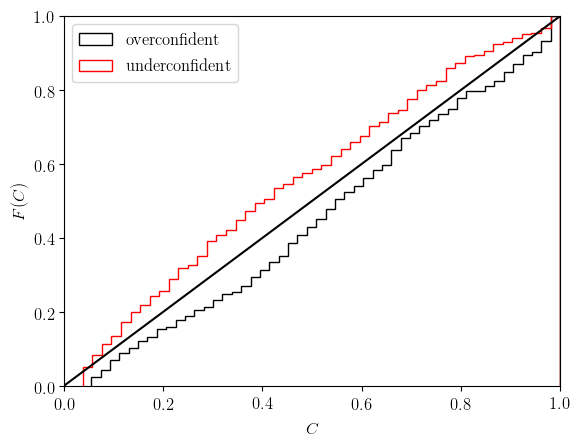}}
 \caption{Credibility analysis examples of \textit{overconfidence}, represented by the black curve below the bisector of the plot $F(C)$ \textit{vs} $C$, and of \textit{underconfidence}, indicated by the red curve above the same bisector.}
 \label{fig:credex}
 \end{figure*}

Finally, we use two additional diagnostics to analyze the \textit{cumulative} performance of the PDFs: the credibility analysis presented in \cite{Wittman2016} and the Probability Integral Transform (hereafter PIT), described in \cite{Gneiting2007}.\\
The credibility test should assess if PDFs have the correct \textit{width} or, in other words, it is a test of the \textit{overconfidence} of any method used to calculate the PDFs. In particular, the method is considered overconfident if the produced PDFs result too narrow, i.e. too sharply peaked, underconfident otherwise.
The implementation of the credibility method is straightforward and is reached by computing the threshold credibility $C_{i}$ for the \textit{i}-th galaxy with
\begin{equation}\label{eq:eqWitt}
C_{i} = \sum_{z\in p_{i} \geq p_{i} (z_{spec,i}) } { p_{i}(z)}
\end{equation}
where $p_{i}$ is the normalized PDF for the \textit{i}-th galaxy. The credibility is then tested by calculating the cumulative distribution \textit{F(C)}, which should be equal to \textit{C}. \textit{F(C)} is a q-q plot, (a typical quantile-quantile plot used to compare two distributions), in which \textit{F} is expected to match \textit{C}, i.e. it follows the bisector in the \textit{F} and \textit{C} ranges equal to [0,1]. Therefore, the \textit{overconfidence} corresponds to \textit{F(c)} falling below the bisector (implying that too few galaxies have zspec with a given credibility interval), otherwise, the \textit{underconfidence} occurs. In both cases, this method indicates the inaccuracy of the error budget \cite{Wittman2016}. \textit{Overconfidence} and \textit{underconfidence} are plotted in Fig.~\ref{fig:credex}.

\begin{figure*}
 \centering
 {\includegraphics[width=0.49 \textwidth]{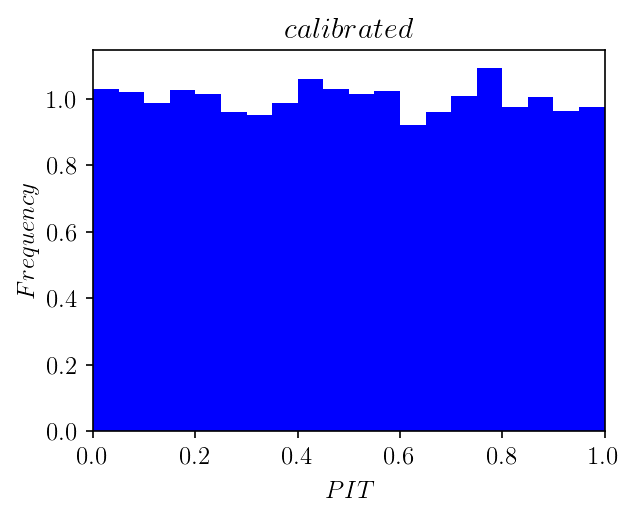}}
 {\includegraphics[width=0.49 \textwidth]{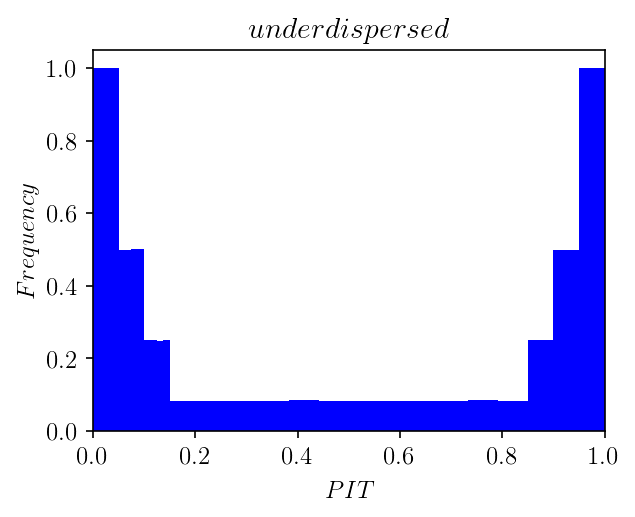}}
 \caption{Examples of well-calibrated (left) and underdispersed U-shaped PIT (right). The overdispersion is simply the opposite of the right panel, with a very broad PDF distribution, having a high peak in the centre of the diagram.}
\label{fig:pitex}
\end{figure*}

The PIT analysis measures how consistent are the predicted zphot and the true redshift (zspec) distributions, by calculating the histogram for the following probabilities:
\begin{equation}\label{eq:PIT}
p_{i}=F_{i}(x_{i})
\end{equation}
$F_{i}$ in Eq.~\ref{eq:PIT} is the cumulative distribution function (CDF) of the i-th object and $x_{i}=zspec_{i}$.
The closer the histogram is to a uniform distribution, the better is the calibration between zphot and zspec distributions. A strongly U-shaped PIT histogram denotes a highly \textit{underdispersive} character of the zphot distribution. \\
The visual inspection of a PIT can, therefore, shed light on the consistency between the zspec and zphot distributions. In particular, if the PDFs are too broad, then the relative PIT histogram appears overdispersed, that is with a peak in the centre of the histogram itself. In contrast, if the PDFs are too narrow, then the PIT is U-shaped and it results underdispersed. Finally, only when the widths of the PDFs agree with the discrepancies between zphot and zspec, then a uniformly distributed PIT histogram is produced. In Fig.~\ref{fig:pitex} an example of well-calibrated and \textit{underdispersed} PIT histograms is shown.

Credibility and PIT tests are complementary since both the underdispersion and the overconﬁdence are related to the narrowness of the PDFs. The narrower the PDFs are, the more the PIT histogram is underdispersed and the results of credibility are overconﬁdent.

\section{Results}
\label{sec:results}

The initial test dataset was composed by $40,796$ zphot estimates and relative individual PDFs. Among these, we have $970$ outlier sources ($2.4\%$), and $39,826$ non-outliers ($97.6\%$): outliers were singled out as explained in Sec.~\ref{sec:method}. We then proceeded with the visual inspection of the individual PDFs, i.e. their width, number of bins, the height of the maximum peak. To do so, we first derived a set of statistical descriptors, such as mean, standard deviation and the minimum and maximum values of the PDF shape properties, dividing the sample into outliers and non-outliers. These values are given in Tables~\ref{tab:pdffeatstatOUT} and ~\ref{tab:pdffeatstatNOOUT}, respectively. By comparing the mean values of such descriptors, the expected differences for the two populations become apparent: outliers have wider PDFs, with a higher number of bins (intervals of amplitude $\Delta z$=0.02), in which the PDF is not null and lower peaks with respect to those for non-outliers samples.

\begin{table}
 \centering
 \caption{ Statistics of the three descriptors of the PDF shape: width, number of bins, and maximum peak height, defined in Sec.~\ref{sec:method}, for the outliers in the test set. }
 \begin{tabular}{|c|c|c|c|c|}
 \hline
 {\bf PDF feature} & {\bf mean} & {\bf $\sigma$} & {\bf Min} & {\bf Max} \\ \hline
 $\textit{PdfWidth}$ & $3.94$\phantom{AA}	& $2.33$ & $0.1$ & $7.02$ \phantom{A}\\
 $\textit{PdfNBins}$ & $124.90$\phantom{AA}	& $71.74$ & $5$ & $277$ \phantom{A}\\
 $\textit{PdfPeakHeight}$ & $0.067$\phantom{AA}	& $0.072$ & $0.011$ & $0.66$ \phantom{A}\\ \hline
 \end{tabular}
 \label{tab:pdffeatstatOUT}
\end{table}

\begin{table}
 \centering
 \caption{ Statistics of the three descriptors of the PDF shape: width, number of bins, and maximum peak height, defined in Sec.~\ref{sec:method}, for the non-outliers in the test set.}
 \begin{tabular}{|c|c|c|c|c|}
 \hline
 {\bf PDF feature} & {\bf mean} & {\bf $\sigma$} & {\bf Min} & {\bf Max} \\ \hline
 $\textit{PdfWidth}$ & $1.23$\phantom{AA}	& $1.82$ & $0.040$ & $7.02$ \phantom{A}\\
 $\textit{PdfNBins}$ & $25.01$\phantom{AA}	& $29.91$ & $2$ & $305$ \phantom{A}\\
 $\textit{PdfPeakHeight}$ & $0.28$\phantom{AA}	& $0.15$ & $0.012$ & $0.99$ \phantom{A}\\ \hline
 \end{tabular}
 \label{tab:pdffeatstatNOOUT}
\end{table}


\noindent For the test set data, in figures~\ref{fig:widthhist}, \ref{fig:heightVSwidth} and \ref{fig:BinsVSwidth} we plot, respectively: \textit{(i)} the histogram of the \textit{pdfWidth} distribution, \textit{(ii)} the scatter plot of \textit{PdfPeakHeight} against \textit{PdfWidth}, and \textit{(iii)} \textit{PdfNBins} versus \textit{PdfWidth}, distinguishing outliers and non-outliers~populations.\\

The inspection of these plots led us to define four data sets: 
\begin{itemize}
\item In Fig.~\ref{fig:widthhist}, we can see that a cut of objects with \textit{PdfWidth} $>4$ can remove a fraction of outliers $\simeq 1.1\%$ from the test sample. We, therefore, define  a first  (\textit{Cut-1})  data set of objects with reliable PDF widths, using the condition:
\begin{equation}\label{eq:widthcond}
pdfWidth<4
\end{equation}
with which we come out with $36,170$ sources (of which $518$ are outliers, and $35,652$ non-outliers, respectively, $1.4\%$ and $98.6\%$ of the total number of objects in the sample).

\begin{figure*}
 \centering
 {\includegraphics[width=0.8 \textwidth]{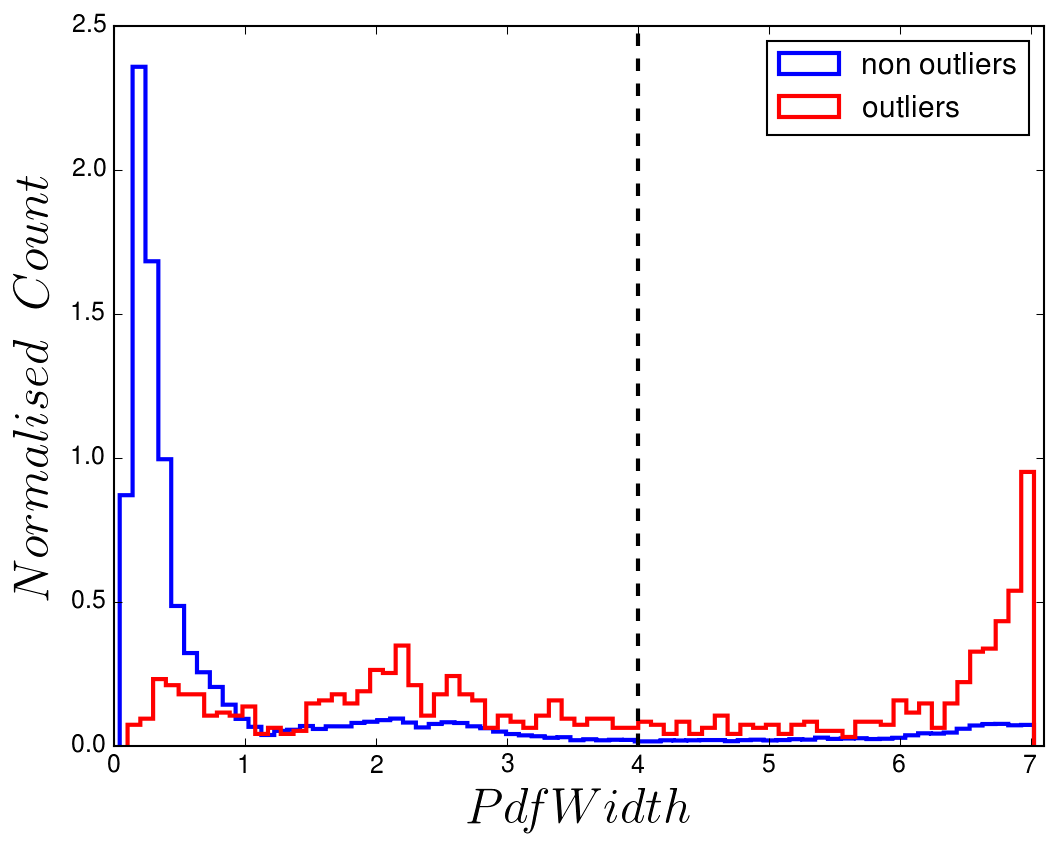}}
 \caption{\textit{PdfWidth} normalized counts for outliers and non-outliers populations. Dashed vertical line: \textit{PdfWidth} value equal to $4$, identified as threshold for clipping outliers that populate the region with widths larger than $4$ (\textit{Cut-1}).}
 \label{fig:widthhist}
 \end{figure*}
 \item In Fig.~\ref{fig:heightVSwidth}, we show the \textit{PdfPeakHeight} versus the \textit{PdfWidth}. Outliers lay at the bottom of the scatter plot, thus allowing to define a second data set (\textit{Cut-2}) via the condition:
\begin{equation}\label{eq:heightcond}
PdfPeakHeight>0.09
\end{equation}
The \textit{Cut-2} data set contains a total of $38,107$ sources: $226$ outliers ($0.6\%$), and $37,881$ non-outliers ($99.4\%$).\\

 \begin{figure*}
 \centering
 {\includegraphics[width=0.8 \textwidth]{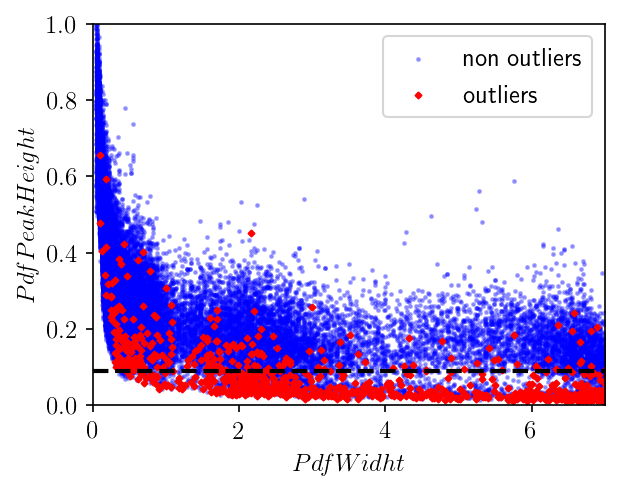}}
 \caption{Scatter plot of \textit{PdfPeakHeight} versus \textit{PdfWidth}. The dashed horizontal line indicates the \textit{PdfPeakHeight} value, equal to $0.09$, identified as threshold for removing outliers laying under the line (see \textit{Cut-2} in the text).}
 \label{fig:heightVSwidth}
 \end{figure*}
 
 \item In Fig.~\ref{fig:BinsVSwidth}, we show the distribution of the descriptors \textit{PdfBins} vs \textit{PdfWidth}. As expected, the majority of the outliers rests in the region with higher values of PDF width and number of bins. This allows defining a third data set \textit{Cut-3}, by rejecting objects with \textit{PdfBins} $>150$. This third data set consists of $39,905$ sources, of which $591$ are outliers ($1,5\%$) and $39,314$ non-outliers ($98.5\%$). \\
 \begin{figure*}
 \centering
 {\includegraphics[width=0.8 \textwidth]{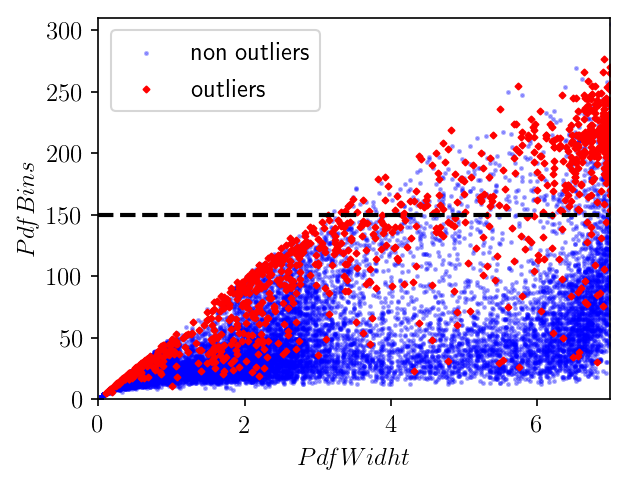}}
 \caption{Scatter plot \textit{PdfNBins} versus \textit{PdfWidth}. The dashed horizontal line identifies the value of \textit{PdfBins} equal to $150$, useful to clip the outliers populating the region above this threshold (see \textit{Cut-3} in the text).}
 \label{fig:BinsVSwidth}
 \end{figure*}
\item Finally, we derived a fourth data set (\textit{Cut-4}) via the combination of \textit{Cut-1} and \textit{Cut-2}. This last data set contains $34,802$ sources, of which $196$ ($0.6\%$) are outliers and $34,606$ ($99.4\%$) non-outliers.\\
\end{itemize}

Some additional tests showed that making more severe cuts would result in an uncomfortable loss in completeness. For instance, by producing an additional data set \textit{Cut-5}, selecting objects with \textit{PdfWidth} $<1$, having less than $150$ bins and a maximum peak height of at least $0.15$, we reduced the test set by $34\%$ ($27,297$ out of $40,796$ sources). 

\subsection{zphot and stacked PDF statistics}
\label{sec:results1}
The results in terms of both  zphot statistics and of \textit{cumulative} PDF performance are reported in Table~\ref{tab:stackedstat}, for the test set and the four data sets corresponding to the different cuts. As we can see, the NMAD statistics is not different for the four adopted cuts, and there is only a slight improvement with respect to the whole data set. It has to be noted, however, that all cuts prove quite effective in reducing the fraction of outliers, with an improvement  of $39.8\%$, $75.0\%$, $37.7\%$, and $76.3\%$ (for \textit{Cut-1}, \textit{Cut-2}, \textit{Cut-3}, and \textit{Cut-4}, respectively) with respect to the test set.\\

\noindent In the case of the data set extracted from \textit{Cut-5}, we were left with a fraction of outliers $\simeq 0.2\%$. The $NMAD$ is reduced to $0.013$, and the fractions of residuals for the stacked PDF, $f_{0.05}$ and, $f_{0.15}$, increase to, respectively, $85.9\%$ and $99.3\%$. This was expected, since the role played by $NMAD$ and fraction of outliers for zphot point estimates, is analogous to the one of, respectively, $f_{0.05}$ and $f_{0.15}$ for the PDF.

\begin{table}
 \centering
 \caption{Statistics of the zphot and stacked PDFs for the whole test set sample and the four pruned data sets performed.}
 \begin{tabular}{|c|c|c|c|c|c|}
 \hline
 {\bf Estimator}	 & {\bf test set} & {\bf cut 1} & {\bf cut 2} & {\bf cut 3} & {\bf cut 4}\\ \hline
$bias$ & $-0.003$\phantom{AA} & $-0.004$ & $-0.002$ & $-0.003$ & $-0.002$ \phantom{A}\\
$NMAD$ & $0.016$\phantom{AA} & $0.015$ & $0.015$ & $0.015$ & $0.015$ \phantom{A}\\
$outliers$ & $2.4\%$\phantom{AA}	& $1.4\%$ & $0.6\%$ & $1.5\%$ & $0.6\%$ \phantom{A}\\
 $f_{0.05}$ & $76.1\%$\phantom{AA}	& $79.2\%$ & $79.8\%$ & $77.5\%$ & $81.2\%$ \phantom{A}\\
 $f_{0.15}$ & $94.6\%$\phantom{AA}	& $96.6\%$ & $97.4\%$ & $95.9\%$ & $98.0\%$ \phantom{A}\\
 $\langle{\Delta z}\rangle$ & $-0.025$\phantom{AA}	& $-0.016$ & $-0.018$ & $-0.008$ & $-0.013$ \phantom{A}\\ \hline
 \end{tabular}
 \label{tab:stackedstat}
\end{table}

\noindent Another estimator that allows quantifying the reliability of the estimated PDF is the \textit{zspecClass} flag as defined in Sec.~\ref{sec:method}. 
The results for \textit{zspecClass} are reported in Table~\ref{tab:zspecclass}. As it could be expected, the best results in terms of fractions of \textit{zspecClass} equal to $0$ and $1$, occur for the data sets \textit{Cut-2} and \textit{Cut-4}, which include the best scores in terms of zphot point estimates and of \textit{cumulative} PDF performances (Table~\ref{tab:stackedstat}). \\
For the \textit{Cut-5} data set, we achieve for classes $0$ and $1$ the scores of $31.0\%$ and $42.8\%$ respectively, and a smaller fraction of objects of class $3$ (only $0.1\%$) with respect to the other data sets. 
Classes $3$ and $4$ quantify the number of objects falling outside the PDF. The distinction between the two classes gives the supplementary information about how far from the PDFs is their zspec.

\begin{figure*}
 \centering
 {\includegraphics[width=0.49 \textwidth]{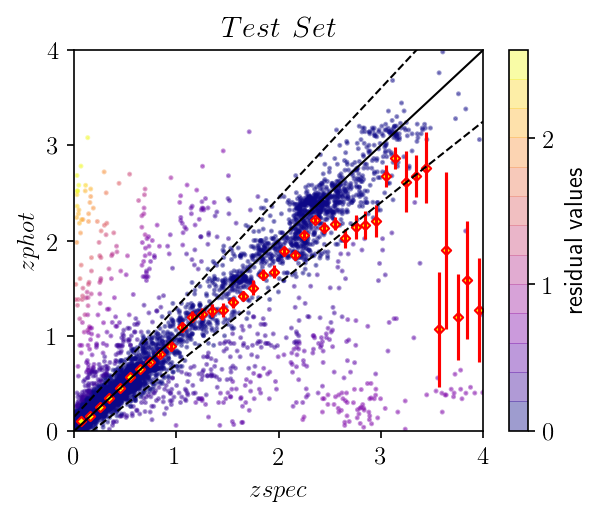}}
 {\includegraphics[width=0.49 \textwidth]{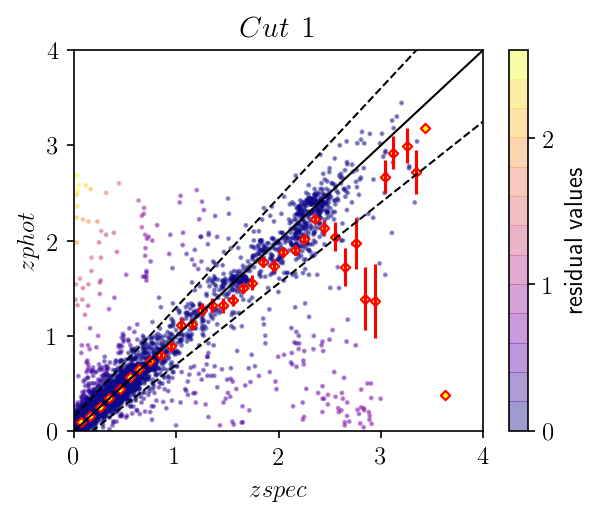}}
 {\includegraphics[width=0.49 \textwidth]{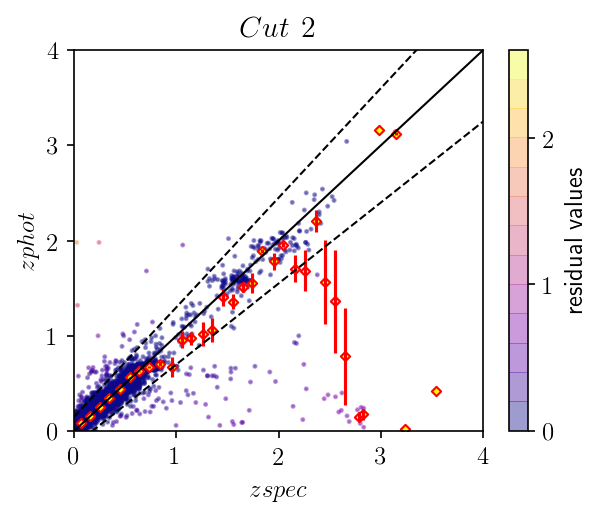}}
 {\includegraphics[width=0.49 \textwidth]{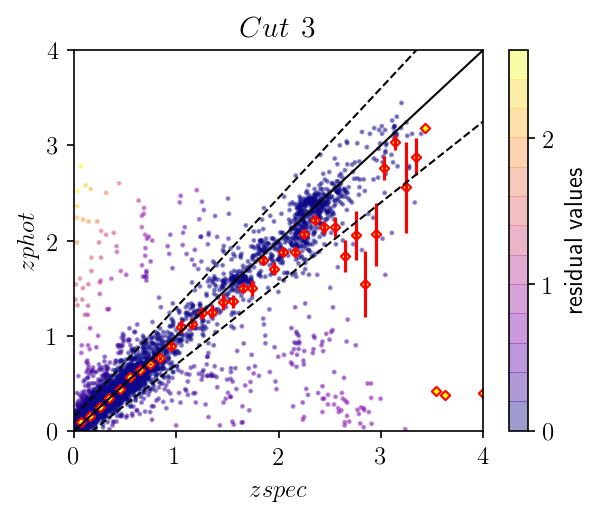}}
 {\includegraphics[width=0.49 \textwidth]{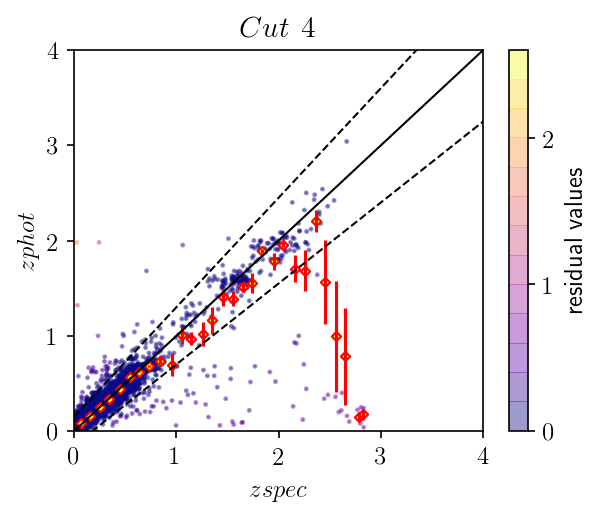}}
 \caption{Scatter plots of photometric redshift \textit{best-estimates} as a function of spectroscopic redshifts. Red diamonds: mean of the zphot \textit{best-estimates} in $40$ evenly spaced zspec bins. Red bars: standard deviation of zphot values populating the bins. Outliers and non-outliers are identified by the colour bar, showing the absolute values of the residuals (see Eq.~\ref{equationDelta}).}
\label{fig:scatterplot}
\end{figure*}

\begin{table}
\centering
\caption{\textit{zspecClass} fractions for the whole test set and the four cuts.}.
\label{tab:zspecclass}
 \begin{tabular}{|c|rr|rr|rr|rr|rr|}
 \hline
{\bf zspecClass} & \multicolumn{2}{c}{\bf test set} & \multicolumn{2}{c}{\bf cut 1} & \multicolumn{2}{c}{\bf cut 2} & \multicolumn{2}{c}{\bf cut 3} & \multicolumn{2}{c}{\bf cut 4}\\ \hline
 0 & $10652$ & $(26.1\%)$ & $9930$& $(27.5\%)$ & $10535$& $(27.6\%)$ & $10631$ &$(26.6\%)$ & $9834$& $(28.3\%)$\\
 1 & $15476$ &$(37.9\%)$ & $14224$ &$(39.3\%)$ & $15214$& $(39.9\%)$ & $15430$ &$(38.7\%)$ & $14084$& $(40.5\%)$ \\
 2 & $13893$ & $(34.0\%)$ & $11353$ &$(31.4\%)$ & $11727$& $(30.8\%)$ & $13115$& $(32.9\%)$ & $10271$& $(29.5\%)$ \\
 3 & $156$ &$(0.4\%)$ & $90$ &$(0.2\%)$ & $79$& $(0.2\%)$ & $115$ & $(0.3\%)$ & $64$& $(0.2\%)$ \\
 4 & $619$& $(1.5\%)$ & $600$& $(1.7\%) $ & $552$& $(1.4\%)$ & $614$ &$(1.5\%)$ & $540$& $(1.6\%)$\\ \hline
\end{tabular}
\end{table}
\noindent In Fig.~\ref{fig:scatterplot} we present the scatter plots of the zphot \textit{best-estimates} as a function of the spectroscopic redshifts, for the test set and the four probed cut data sets. The mean and standard deviation of zphot are also plotted, in $40$ evenly spaced zspec bins in a whole range of $[0, 4.0]$. 
Not all  bins are populated, due to the reduction of the amount of samples resulting from the application of the rejection criteria, and the $\sigma$ value in each bin increases in under-sampled bins. \\

 It is interesting to notice the similar trends for data sets deriving from cuts $1$ and $3$, and cuts $2$ and $4$. This is expected for cut data sets $1$ and $3$ since, as we mentioned, PDF width and the number of bins in which PDF differ from $0$, are highly correlated.
In the case of cut data sets $2$ and $4$, being the cut data set $4$ obtained by the joint application of cuts $1$ and $2$ conditions, the similar performance can shed light on the cut which drives the statistical performance outcome. \\
Of course, we should favour those rejection criteria that, leading to similar statistical performance, remove a smaller number of sources from the original data set. In other words, we should adopt rejection criteria corresponding to the best trade-off between precision and completeness. 

Among the tested cuts, the best results in terms of precision and completeness are achieved for the data set \textit{Cut-2}, which while showing comparable results (cf.~ tables~\ref{tab:stackedstat} and \ref{tab:zspecclass}) to data set \textit{Cut-4}, contains $\simeq 9\%$ more sources.

\begin{figure*}
 \centering
 {\includegraphics[width=0.9\textwidth]{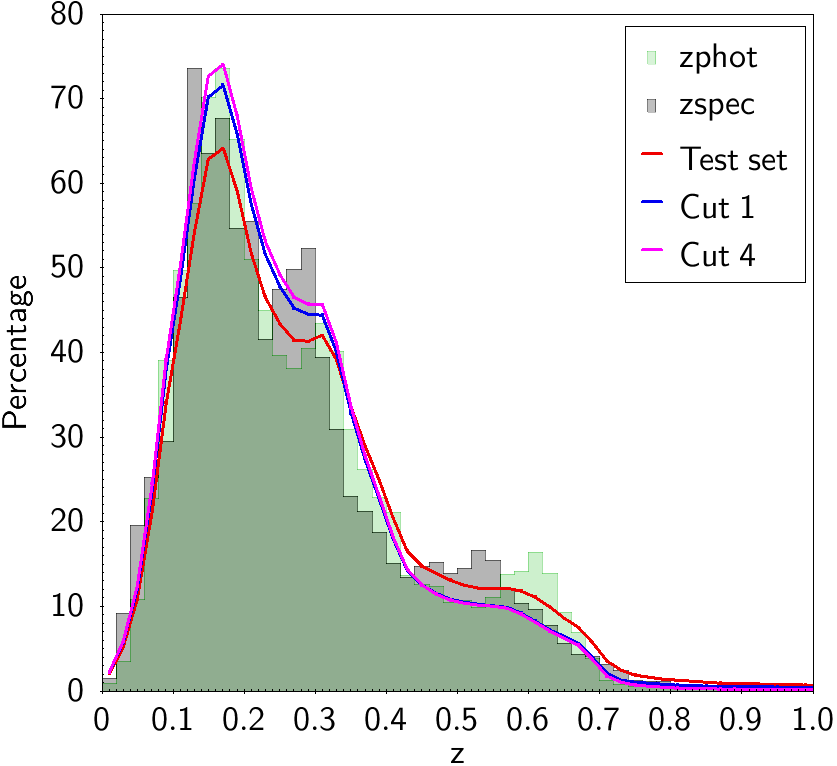}
 }
 \caption{Superposition of the stacked PDF (percentage) of the test set (red), for the \textit{cut 1} (blue) and \textit{cut 4} (magenta) data sets, and the zphot \textit{best-estimates} (in green) distributions obtained by METAPHOR applied to the zspec distribution (in black), for the test set, limited to $z=1$, due to the very few objects over such value.}
 \label{fig:stackedwithCuts}
 \end{figure*}

Finally, in Fig.~\ref{fig:stackedwithCuts}, we show the stacked PDF for the test set and two cut data sets (\textit{Cut-1} and \textit{Cut-4}), along with the spectro-photometric redshift distributions of the test set. Note that the redshift range has been cut at $z=1$, due to the low amount of objects in the test set above such value.\\
Besides the good agreement with the zphot distribution for the stacked PDF of the test set, we can see the effect of the rejection for the clipped data sets, which leads to a lower amount of sources at high redshift, and a more substantial amount at low redshift.

\begin{figure*}
 \centering
 {\includegraphics[width=0.85 \textwidth]{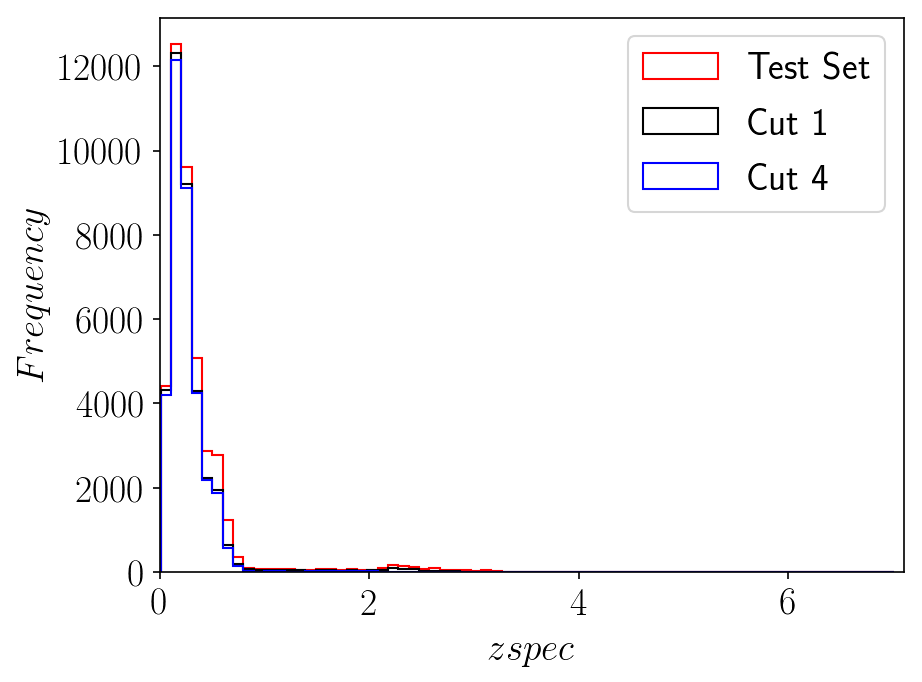}
 }
 {\includegraphics[width=0.85 \textwidth]{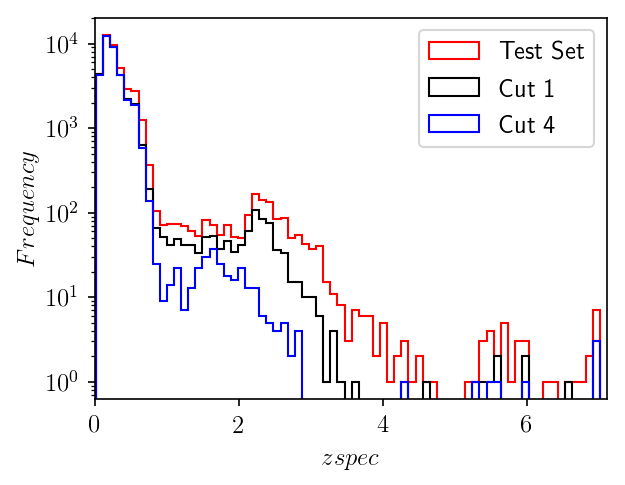}
 }
 \caption{\textit{Top panel:} zspec distribution for the whole test set (red), and the cut data sets \textit{Cut 1} (black) and \textit{Cut 4} (blue). \textit{Bottom panel}: the same of the top panel in a logarithmic scale.}
 \label{fig:distrib_zspec}
 \end{figure*}

This is highlighted in Fig.~\ref{fig:distrib_zspec}, where the distribution of zspec for the whole test set and the two clipped data sets \textit{Cut-1} and \textit{Cut-4} are shown. Moreover, in Fig.~\ref{fig:distrib_zspec_allcuts} in the Appendix, it is reported the zspec distribution for the whole test set against the zspec distributions for the tested four clipped data sets. As we can see more clearly from this figure, rejection is successful in removing outliers at higher redshift. Comparing the distributions for the \textit{Cut-2} and \textit{Cut-4} data sets, we can notice that the range of zspec between $3$ and $4$ is more populated in the case of \textit{Cut-2} data set with respect to \textit{Cut-4} data set. This leads to the conclusion that, as anticipated above, \textit{Cut-2} achieves a better trade-off between completeness and precision with respect to \textit{Cut-4}.
This is a clue of the effectiveness of the rejection in removing most of outliers at high redshift, i.e. in a region of the parameter space where the density of the training points is lower.

\subsection{PIT and credibility analysis} 
\label{sec:results2}
\begin{figure*}
 \centering
 {\includegraphics[width=0.49 \textwidth]{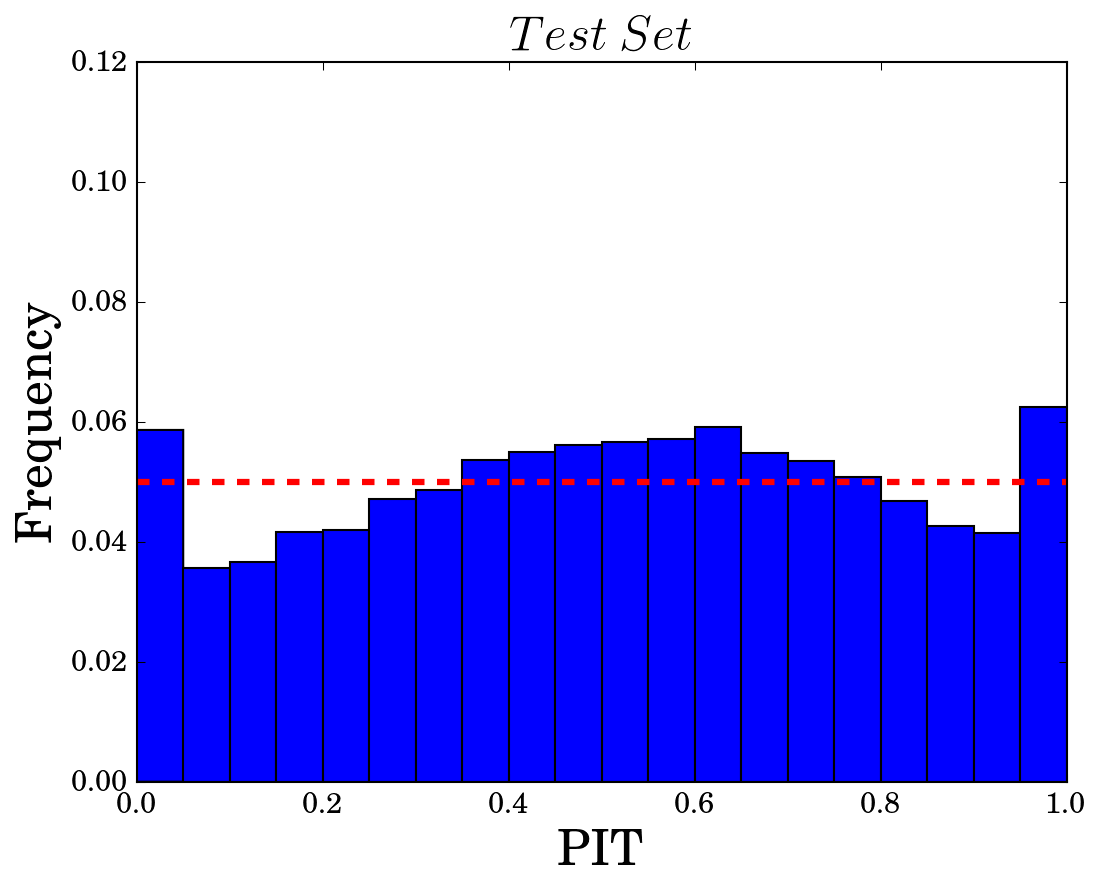}}
 {\includegraphics[width=0.49 \textwidth]{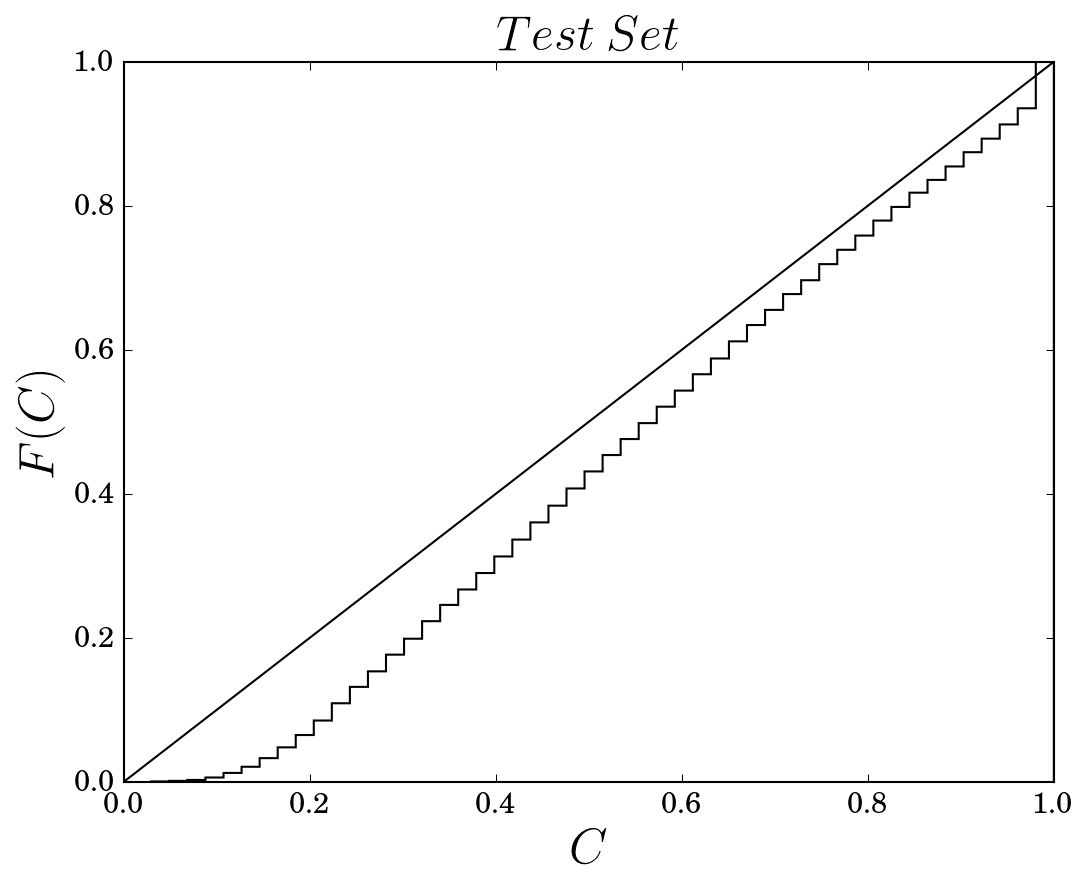}}
 \caption{PIT histogram (left panel). The red dashed line identifies the ideal PIT value, which represents the best calibration between the true zspec and the reconstructed zphot distributions. Credibility analysis (right panel) for the test set. 
The solid black line represents the best credibility for which the two distributions \textit{F(C)} and \textit{C} are indistinguishable (see Sec.~\ref{sec:method}).}
 \label{fig:pitCredTestSet}
 \end{figure*}

PIT and credibility analysis for the test set are shown in Fig.~\ref {fig:pitCredTestSet}. The PIT histogram shows a certain degree of \textit{underdispersion} of the zphot distribution and the credibility plot stresses the \textit{overconfidence} of the PDFs. The complementary information carried by these two \textit{visual} diagnostics (see Sec.~\ref{sec:method}) is therefore confirmed. 

\begin{figure*}
 \centering
 {\includegraphics[width=0.49 \textwidth]{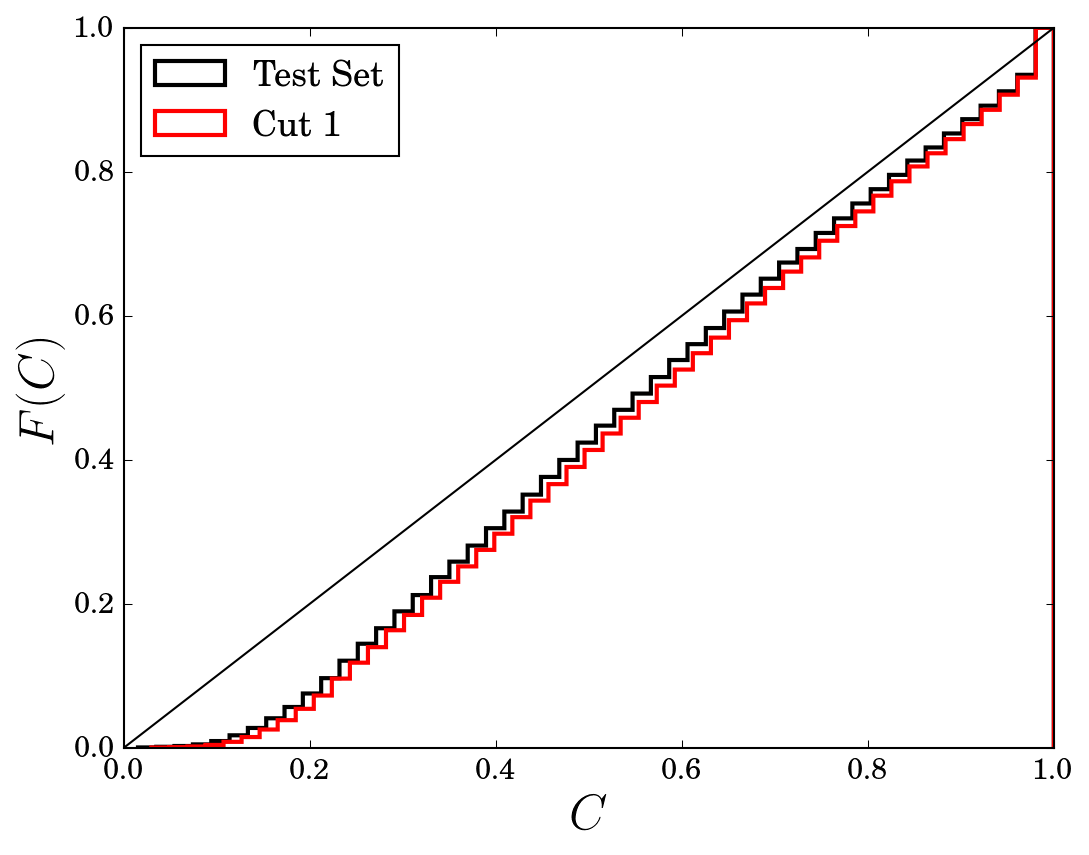}}
 {\includegraphics[width=0.49 \textwidth]{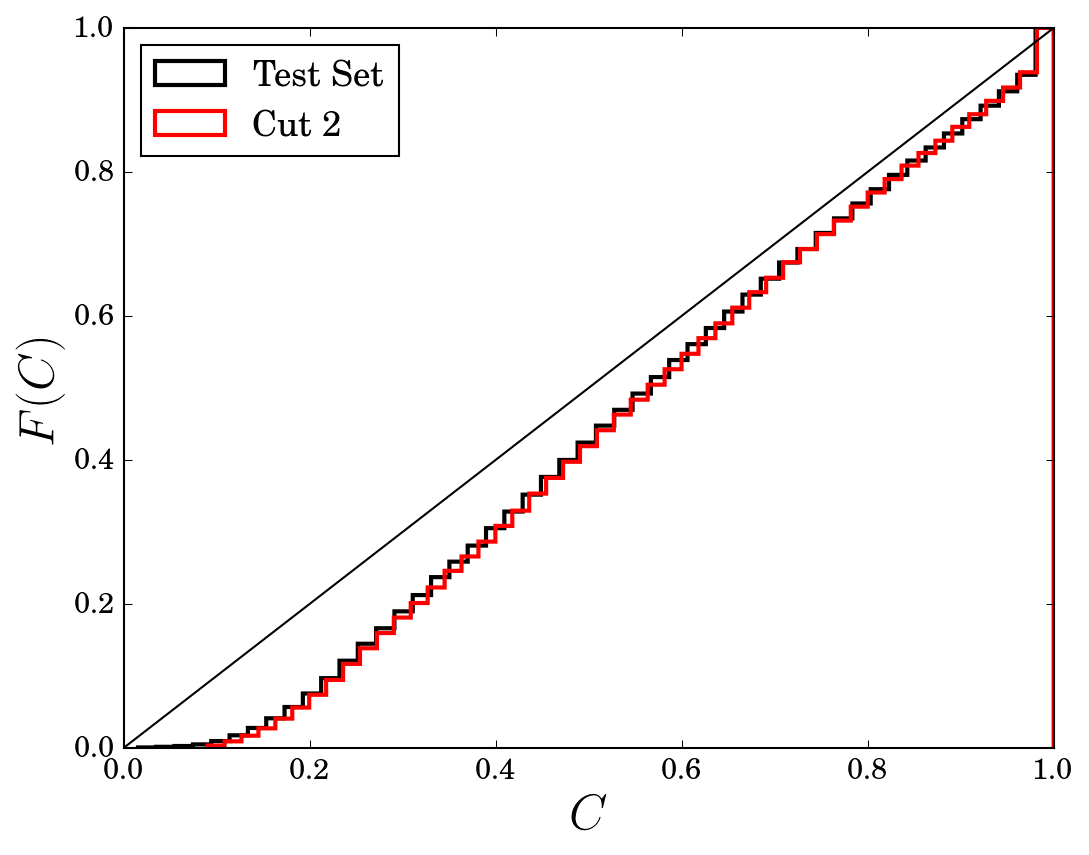}}
 {\includegraphics[width=0.49 \textwidth]{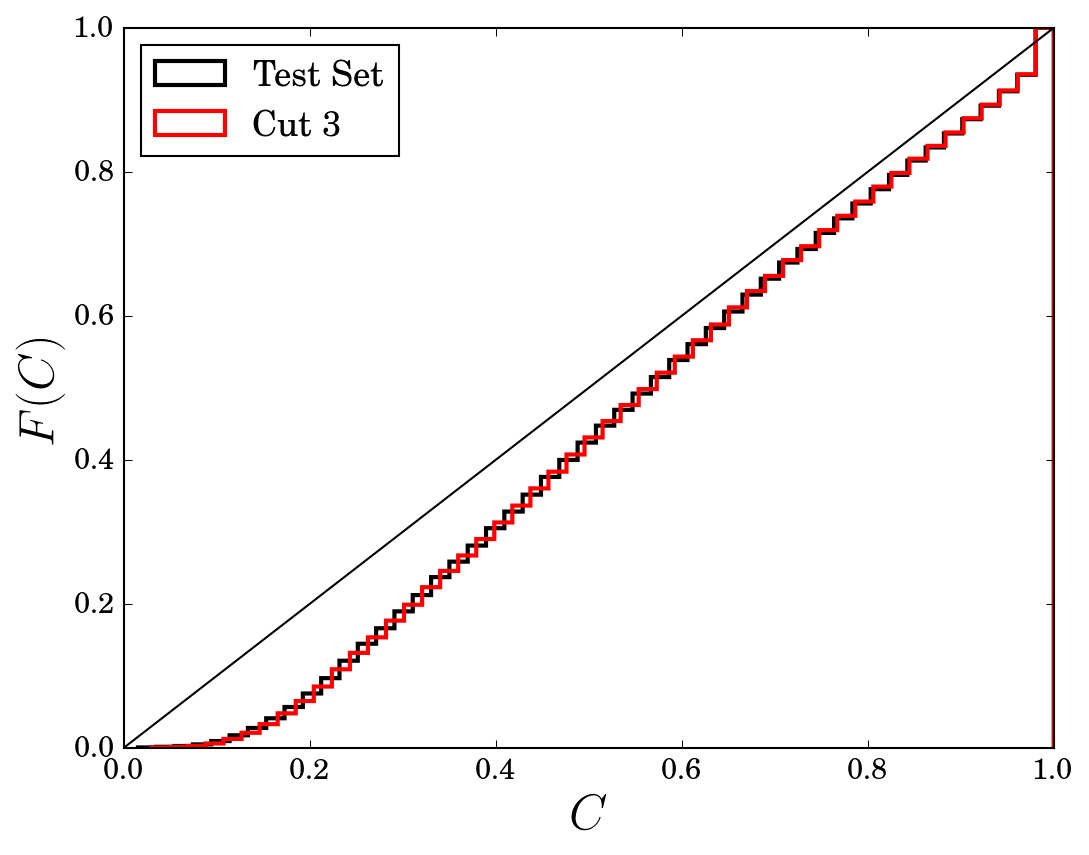}}
 {\includegraphics[width=0.49 \textwidth]{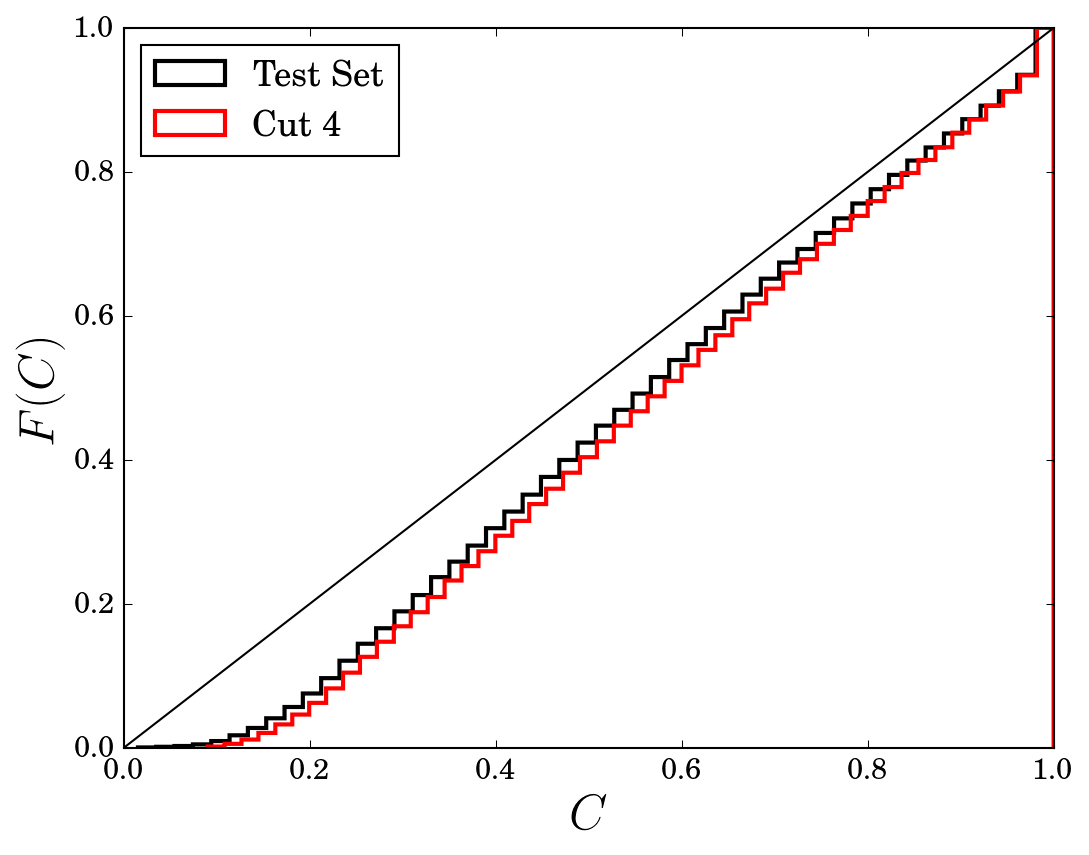}}
 \caption{Credibility analysis for the four tested cut data sets (red) against the whole test set credibility (black). In the top panels: \textit{Cut-1} (left) and \textit{Cut-2} (right). In the bottom panels: \textit{Cut-3} (left) and \textit{Cut-4} (right).}
 \label{fig:Credallcuts}
 \end{figure*}

\begin{figure*}
 \centering
 {\includegraphics[width=0.49 \textwidth]{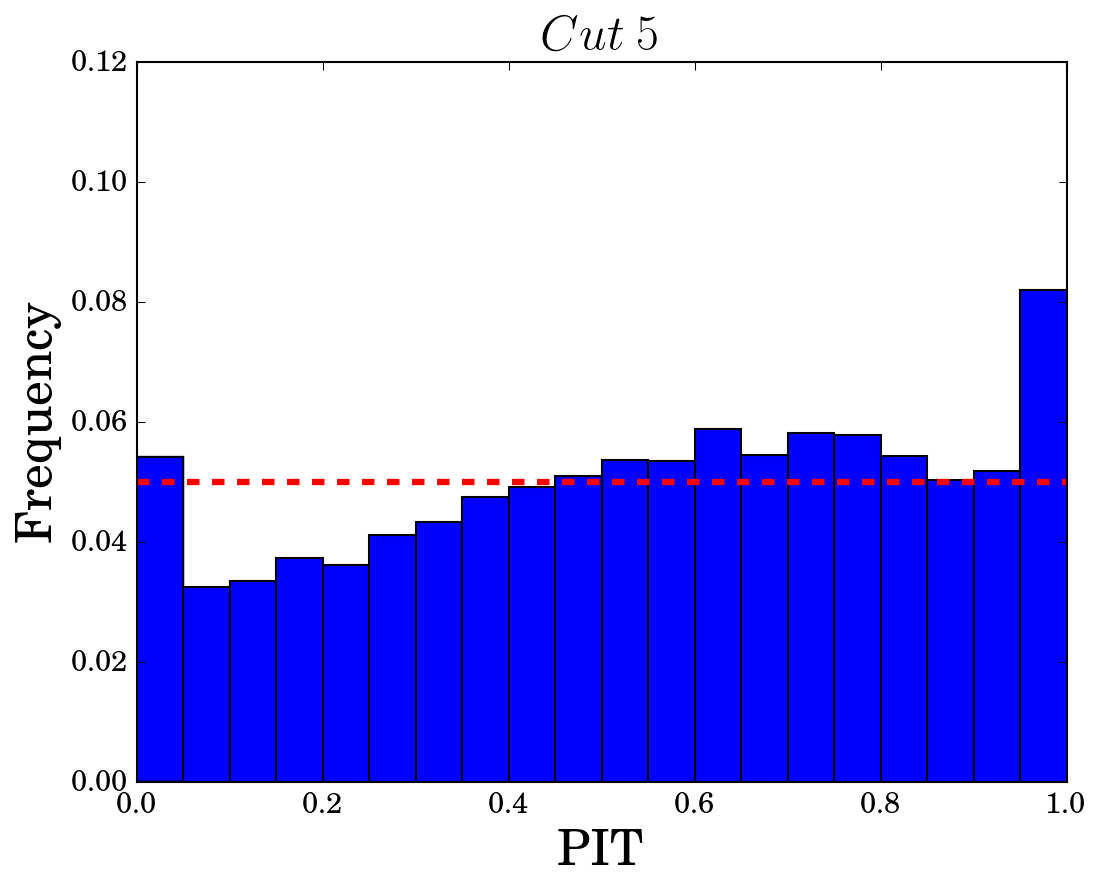}}
 {\includegraphics[width=0.49 \textwidth]{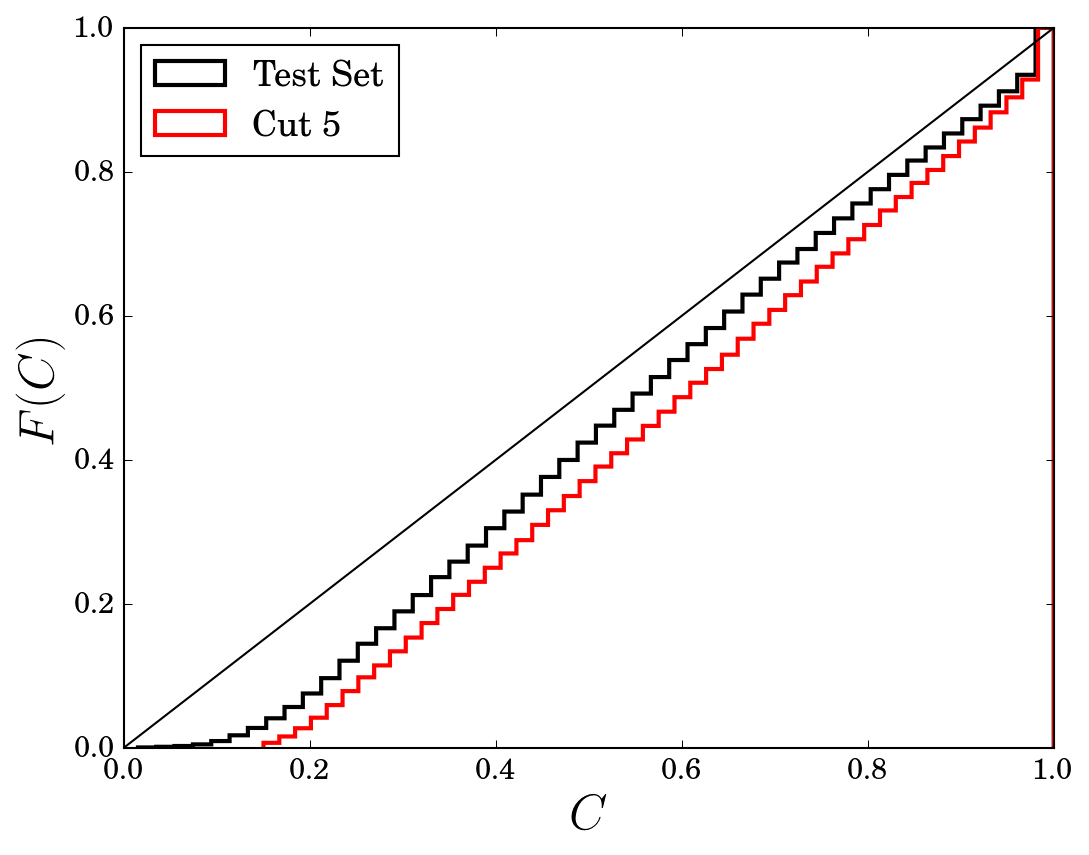}}
 \caption{PIT histogram (left panel) and credibility analysis (right panel) for the \textit{Cut-5} data set, plotted against that of the whole test set.}
 \label{fig:PITcredCut5}
 \end{figure*}

In the plots of figures~\ref{fig:Credallcuts} and ~\ref{fig:PITcredCut5}, we show, respectively, the credibility analysis for the four clipped data sets against the credibility of the test set, and the comparison of PIT and credibility for the data set  \textit{Cut-5}. Data sets obtained from cuts $1$ and $4$ show a slightly higher degree of \textit{overconfidence} with respect to the test set, while cut data sets $2$ and $3$ show an indistinguishable credibility trend. \\
\indent We stress that the PIT histogram fails to reveal differences between the four 
clipped data sets probed, with respect to the test set: for this reason, we do not show the relative plots. However, in the case of \textit{Cut-5}, PIT shows a more significant degree of bias with respect to the test set, whereas the credibility shows a narrower shape, resulting in a larger \textit{overconfidence} with respect to the whole test set. 

\section{Conclusions}
\label{sec:conclusion}
\begin{figure*}
 \centering
 {\includegraphics[width=0.49 \textwidth]{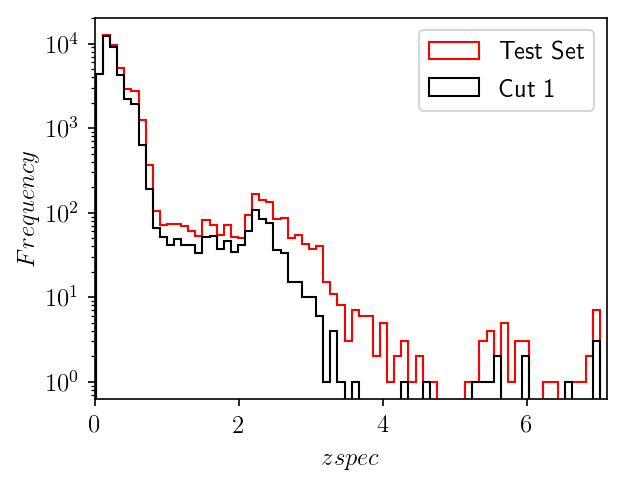}}
 {\includegraphics[width=0.49 \textwidth]{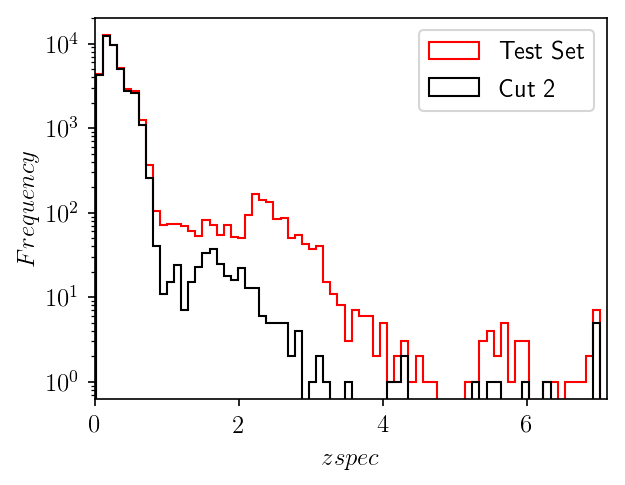}}
 {\includegraphics[width=0.49 \textwidth]{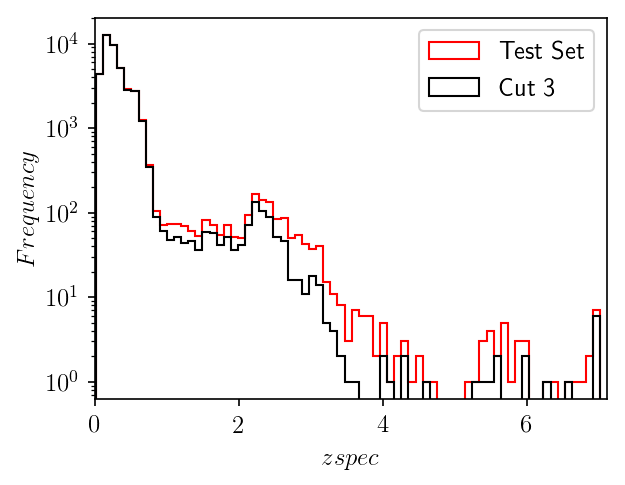}}
 {\includegraphics[width=0.49 \textwidth]{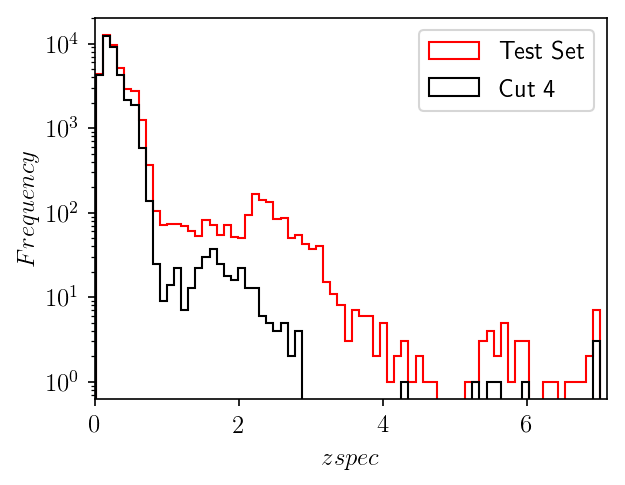}}
  \caption{Zspec distribution for the whole test set (red) and the four tested cut data sets (black), in a logarithmic scale. \textit{Top panels}: \textit{Cut-1} (left) and \textit{Cut-2} (right). \textit{Bottom panels}: \textit{Cut-3} (left) and \textit{Cut-4} (right).}
\label{fig:distrib_zspec_allcuts}
\end{figure*}

In this work, we presented a method for defining low-quality zphot rejection criteria through the characterization of outliers, using the descriptors of the PDF shape (e.g. the width, the value of the maximum peak, etc.). 
The first step was, therefore, to compare the PDF descriptors for the two populations of outliers and non-outliers. Outliers appear to be characterized by wider PDFs with small maximum probability, as well as by a more significant number of bins in which the PDF differs from zero. \\
Zphot outliers tend to populate particular regions of the photometric  parameter space and of the one defined by the PDF characteristics. 
Most outliers populate  the top right part of a plot \textit{PdfNBins} vs \textit{PdfWidth}, where both the quantities are larger. 
Furthermore, the PDFs of the outliers have low maximum peaks, and populate a stripe at low values of \textit{PdfPeakHeight}, in a plane \textit{PdfPeakHeight} vs \textit{PdfWidth}. 
This allows the identification of cuts suitable to remove outliers, thus improving the precision on the clipped data sets. \\
We detailed the results for four different cut data sets obtained by applying rejection through the PDF width, the  height of the maximum peak, and the number of bins in which the PDF is not null for, respectively \textit{Cut-1, 2, 3} data sets. A further clipped data set (\textit{Cut-4)} was created by removing outliers through the application of both the cuts used to generate \textit{Cut-1} and \textit{Cut-2} data sets. The best precision and completeness results were achieved for the data set \textit{Cut-2}. This data set, in fact,  from the one hand, shows comparable results to data set \textit{Cut-4}, in terms of both zphot point estimate and \textit{cumulative} PDF statistical performances. On the other hand, \textit{Cut-2} data set contains $\simeq 9\%$ more sources than \textit{Cut-4}, which mostly populate the spectroscopic region in the range [3,4], as it is visible in Fig.~\ref{fig:distrib_zspec_allcuts}.\\
Finally, we tested many others more strict rejection criteria, all of them leading to a severe loss of completeness with respect to the original data set. We reported for one of these pruned data set (\textit{Cut-5}) the results throughout the Sec.~\ref{sec:results}, also showing the more biased PIT histogram and more \textit{overconfident} credibility diagram with respect to the other four pruned data sets (see Fig.~\ref{fig:PITcredCut5}).\\
Although still not fully automated, the rejection approach is very general in its applicability, since it does not depend on the particular method used to calculate the PDFs. On the other hand, the overall quality of PDFs depends strictly on the particular method used to derive them. This last aspect is not discussed in this paper. However, we deem particularly useful a future comparison of rejections applied to PDFs obtained by different approaches (e.g. SED and ML methods referenced in Sec.~\ref{sec:PDF}). This with the final goal of further increasing the precision of the measurements. \\
As mentioned in Sec.~\ref{sec:intro}, precision and completeness are both relevant quantities for matching the requirements of ongoing as well as future cosmological sky surveys, since the accuracy of the cosmological parameters strongly depends on an optimal trade-off between these two properties. The systematic study and automatic implementation  of rejection can help to improve the precision, keeping a congruous number of non-outliers objects, thus preserving the completeness.

\begin{acknowledgement}
Based on observations made with ESO Telescopes at the La Silla Paranal Observatory under programme IDs 177.A-3016, 177.A-3017, 177.A-3018 and 179.A-2004, and on data products produced by the KiDS consortium. The KiDS production team acknowledges support from: Deutsche Forschungsgemeinschaft, ERC, NOVA and NWO-M grants; Target; the University of Padova, and the University Federico II (Naples). SC acknowledges the financial contribution from FFABR 2017. GL acknowledges partial financial support from the EU ITN SUNDIAL.
MB acknowledges financial contributions from the agreement \textit{ASI/INAF 2018-23-HH.0, Euclid ESA mission - Phase D}. MB and CT acknowledge the \textit{INAF PRIN-SKA 2017 program 1.05.01.88.04}.
\end{acknowledgement}

\section*{Appendix}
\addcontentsline{toc}{section}{Appendix}

\bibliographystyle{spphys}
\bibliography{main}{}

\end{document}